%% file: monrabal.tex
\pdfoutput=1 
\documentclass{JINST}

\usepackage{amsmath,amssymb}
\usepackage[numbers,sort]{natbib}

\title{An homeopathic cure to pure Xenon large diffusion}
\input{defs.tex}

 \author{
C. D. R. Azevedo$^a$, L.M.P. Fernandes$^b$, E. D. C. Freitas$^b$, D. Gonzalez-Diaz$^c$,\thanks{Presenter at LIDINE 2015}\hspace{0.1cm}  F. Monrabal$^d$, C. M. B. Monteiro$^b$, J. M. F. Dos Santos$^b$, J.F.C.A. Veloso$^a$ and J. J Gomez-Cadenas$^e$\\
 \llap{$^a$}Institute of Nanostructures, Nanomodelling and Nanofabrication (i3N), \\
Universidade de Aveiro, Campus de Santiago, 3810-193 Aveiro, Portugal\\
\llap{$^b$}LIBPhys, Departamento de Fisica, Universidade de Coimbra,\\
Rua Larga, 3004-516 Coimbra, Portugal\\
\llap{$^c$}CERN, European Organization for Nuclear Research,\\
1211 Geneva, Switzerland,\\
 \llap{$^d$}University of Texas at Arlington,\\
  TX 76019, USA\\
 \llap{$^e$} Instituto de Fisica Corpuscular (IFIC), \\
CSIC  Universitat de Valencia,\\
  \\
  E-mail: \email{francesc.monrabalcapilla@uta.edu}}

\abstract{The NEXT neutrinoless double beta decay (\bbonu) experiment will use a high-pressure gas electroluminescence-based  TPC to
search for the decay of Xe-136. One of the main advantages of this technology is the possibility to reconstruct the topology of events with energies close to \qbb .

The rejection potential associated to the topology reconstruction is limited by our capacity to properly reconstruct the original path of the electrons in the gas. This reconstruction is limited by different factors that include the geometry of the detector, the density of the sensors in the tracking plane and the separation among them, etc. Ultimately, the resolution is limited by the physics of electron diffusion in the gas.

In this paper we present a series of molecular additives that can be used in Xenon gas at very low partial pressure to reduce both longitudinal and transverse diffusion. We will show the results of different Monte-Carlo simulations of electron transport in the gas mixtures from wich we have extracted the value of some important parameters like diffusion, drift velocity and light yields.

These results show that there is a series of candidates that can reduce diffusion without affecting the energy resolution of the detector and they should be studied experimentally. A comparison with preliminary results from such an ongoing experimental effort is given.

}


\begin{document}

\section{Introduction and Topological signature}
\vspace{-0.3cm}

Neutrinos have mass and mix. This experimental observation, recently awarded with a Nobel prize, has opened new questions about the nature of neutrinos and its position in the Standard Model: What is the absolute mass of the neutrino? How can it be accomodated in the Standard Model? What is the origin of the neutrino mass? Are neutrinos Majorana particles?

Neutrinos are the only chargeless fermions, then they are the only fermions that can have a Majora mass term. If neutrinos are Majorana particles they will be indistinguishable from their antiparticle and it may lead to processes violating lepton number conservation. Such processes can be connected with the matter-antimatter asymmetry in the universe via leptogenesis .The Majorana mass of the neutrino also provides an elegant explanation of the smallness of neutrino mass via the seesaw mechanism (see \cite{GomezCadenas:2011it} and references therein).

The most sensitive method to establish the nature of the neutrinos is the search for neutrinoless double beta decay (\bbonu). This is a hypothetical, very rare nuclear transition in which a nucleus with Z protons decays into a nucleus with Z+2 protons and the same mass number, A, emitting two electrons that carry essentially all the energy released (\Qbb). While the two-neutrino mode of the double beta decay has already been measured in a number of isotopes, the zero-neutrino mode remains unobserved.

The experimental signature of a neutrinoless double beta decay is the emission of two electrons from the same point with a total kinetic energy always equal to \qbb. For this reason, most of the experiments have focused on improving the energy resolution of their detectors in order to reduce the number of background events that fall inside the region of interest. Of course, any extra handle like the ability of topologically separate background from signal or tagging the daughter nucleus will improve the background rejection resulting in better sensitivity to the process.

The NEXT experiment seeks to make a first measurement of \bbonu\ in \Xe\ using a high pressure gas Time Projection Chamber (TPC) with electroluminescent (EL) read-out. It has been designed to provide good energy and spatial resolution to identify separated tracks and increased ionization ('blobs') at their ends. With this aim, it uses two different planes for energy measurement and tracking. Photomultipliers behind the cathode detect the primary scintillation light and allow for energy measurement by detecting the electroluminescence. An array of silicon photomultipliers (SiPMs) behind the anode provides the spatial information for the topological analysis of the events using the EL light. The \bbonu\ search will be carried out using NEXT-100 which will contain $\sim$100 kg of  \Xe\ gas at 15 bar. The first phase of the experiment, called NEW, and deploying 10 kg (and 20\% of the sensors), is currently being commissioned at the Laboratorio Subterr\'aneo de Canfranc. The principle of operation of NEXT-100 and NEW and the necessary know-how has been developed using NEXT-DEMO, a large scale prototype which operated at the Instituto de F\'isica Corpuscular in Valencia (Spain) with $\sim$1.5 kg of natural xenon at a pressure of 10 bar.


The topological signature of a \bbonu\ event in NEXT is determined by the movement of low energy ($\sim$ MeV) electrons in dense gas. The electrons move in a random path due to collisional scattering with gas atoms. Along this path the electron deposits energy at a fairly constant rate until the end of the trajectory where the energy deposition increases significantly in a small region due to the increase in $dE/dx$ known as Bragg peak; we call that region 'blob'.  Those blobs have an energy around 300 keV but with large fluctuations.
The characteristics of a signal event are then a continuous track with two blobs at both endpoints. On the other hand, the background that is mainly produced by high energy gammas from \BI\ and \Tl\ decays, converts in the gas producing typically more than one electron separated in space.



\begin{figure}[h!]
\centering
\includegraphics[width=0.78\textwidth]{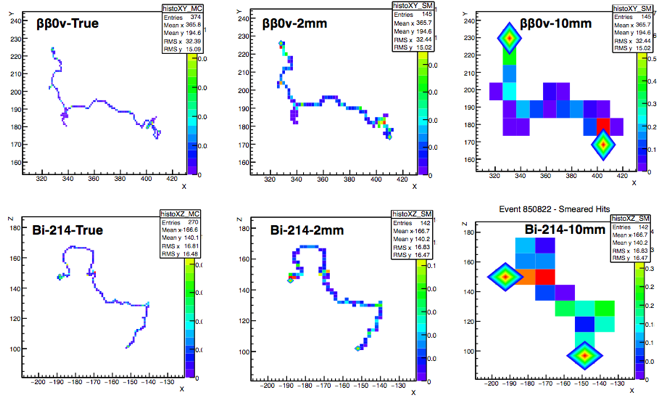}
\caption{Influence of diffusion on the tracking capabilities of our detector. From left to righ we show two simulated events (one for signal and one for background) where we apply different electron diffusion. The central image shows the effect of 2mm diffusion where most of the details that can be seen in the `true' track are still present. The right image shows the track after the diffusion expected for drifting electrons along 1m in pure Xenon has been applied, where most of the details have been lost.} \label{fig:topology}
\end{figure}

In a real detector, the capabilities of identifying the blobs and separating different energy depositions in the detector is limited by the intrinsic position resolution of the detector. Figure \ref{fig:topology} shows how the increase in diffusion transforms the topological signature. When the diffusion is at very low levels (2mm) the electron path is clearly visible, identification of track end-points and blobs is done with high efficiency and low contamination. When the diffusion for pure Xenon along 1m drift is included, the finer details of the track are lost and the topological signature of the events has a weaker effect.  At a different level, algorithms that can follow the real path of the electron in the gas will help identifying the end-points of the electron path improving our blob identification, but the behaviour of such algorithms is also closely related with the position resolution. A more detailed explanation of the topological signature and a description and experimental results of the first efforts of the NEXT Collaboration in that direction can be found in \citep{Ferrario:2015kta}.

\vspace{-0.3cm}

\section{Limits of tracking in pure xenon}
\vspace{-0.3cm}


The spatial resolution in a pure xenon gaseous detector is limited by different factors that can be separated in two different groups: instrumental and physical. In this section we will briefly describe them and evaluate their effect on the current NEXT detector.


The transverse spatial resolution is limited by the pitch of the SiPMs sensors. The minimum resolution we have in this case is the pitch over square root of 12. In NEXT detector it gives a resolution of 3.5 mm that can easily be reduced using simple reconstruction algorithms like barycenter.
The longitudinal resolution is limited by the width of the EL region. In our case it gives a resolution of 1-1.5 mm.
We can see that with the current configuration the instrumental limit is on the order of a couple of milimiters, and we would like to reach this limit. While reducing the pitch of the SiPMs seems feasible, reducing the EL gap in a large detector is a complicated issue that limits the longitudinal resolution.


On the other hand, the diffusion of electrons in pure Xenon adds an effect of the order of $\sim$ 10 mm to the transverse resolution and $\sim$ 4-5 mm to the longitudinal after a meter of drift. Thus, the largest effect affecting our resolution comes from the diffusion of the secondary electrons.

\vspace{-0.3cm}

\section{The effect of additives}
\vspace{-0.3cm}

In this section we show the results of the Magboltz simulations for different concentrations of \cod, \chf\  and \cff, gases that are common in TPCs, and whose elementary cross-sections are well known. The results we are interested in are the diffusion (both transversal and longitudinal), the drift velocity and how the concentrations studied affect the amount of light produced in the electroluminescence region and the energy resolution at different electric fields.

\begin{figure}[h!]
\centering
\includegraphics[width=0.41\textwidth]{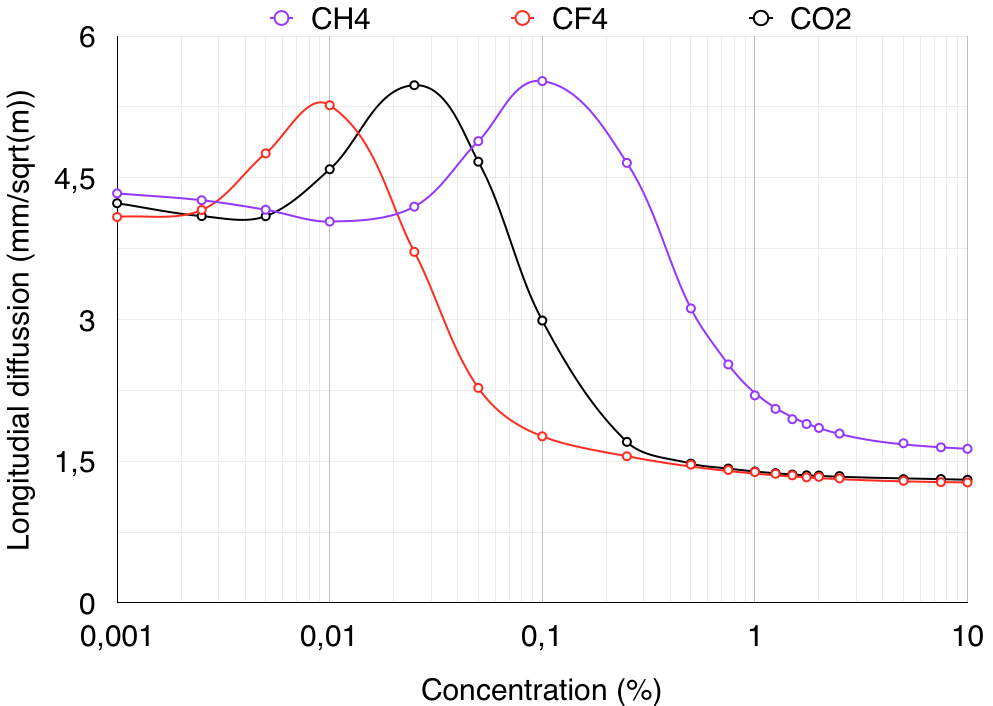}\hspace{1cm}
\includegraphics[width=0.41\textwidth]{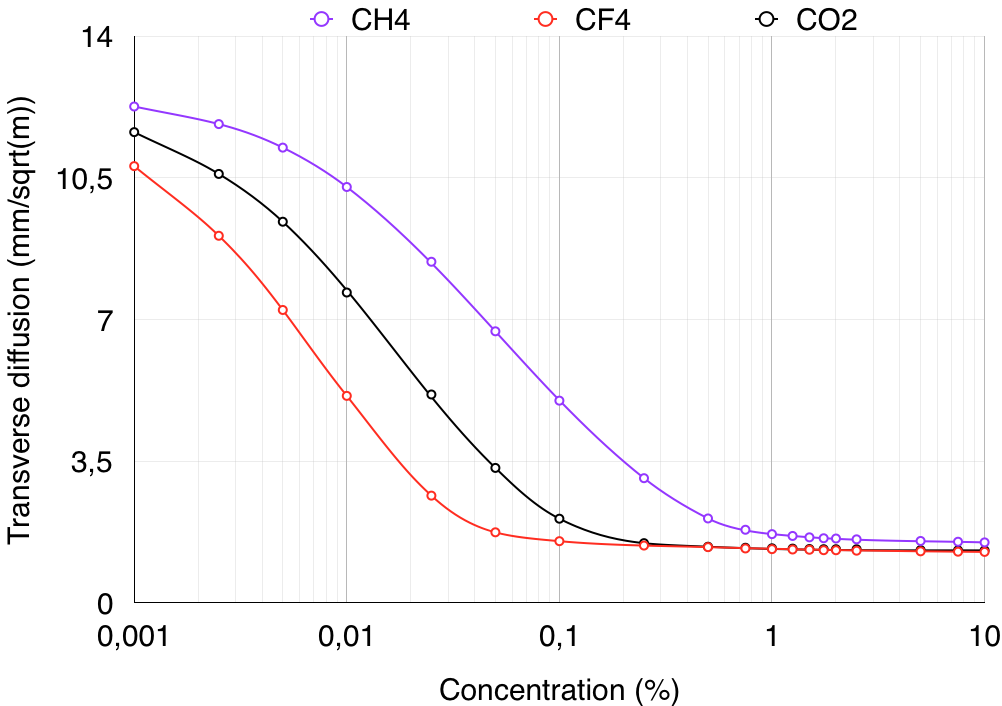}
\caption{Longitudinal and transverse diffusion of electrons in gas for different concentrations of \cod\ , \chf\ and \cff\ for a drift field of 30V/cm/bar and at 10 bar pressure. } \label{fig:ld}
\end{figure}

Figure \ref{fig:ld} shows the results of the simulation for Xenon at 10 bar and different concentrations of the molecules described before. The left image shows the results for the longitudinal diffusion for a typical drift field in NEXT of 30V/cm/bar while the right image shows the result for transverse diffusion. They show that concentrations below the percent level may be enough in the case of \cod\ and \cff\ for reducing the diffusion by a factor 3. This will result in a diffusion on the order of 2 mm for

$1\mathrm{mm}/\sqrt{\mathrm{m}}$ and a transverse diffusion of $\sim 3.5\mathrm{mm}/\sqrt{\mathrm{m}}$. At this level the diffusion will be on the order of the error introduced by the instrumentation.

\vspace{-0.3cm}

\subsection{Cooling electrons in gas}

Energy loss by electrons through elastic collisions with atoms is extremely inefficient.
This determines the large diffusion that is characteristic of noble gases, in particular in the range of reduced
fields of interest for $\beta\beta0$ experiments in $^{136}$Xe ($E\lesssim100$ V/cm/bar). Energy losses can be 
largely facilitated with the addition of even simple molecules, provided new molecular degrees of freedom 
(e.g. vibrations) are made available for energy transfer. In this case, for relatively low fields,
electron energy distributions become just mildly non-thermal and tend to build up around the energy 
of the first vibrational level ($\sim 0.1$eV). Due to this, as seen in Fig. \ref{fig:ld}, diffusion is expected
to be largely reduced even in the presence of very minute concentrations of molecular additives. The concentration of the mixtures we are considering is always below the percent level, which implies the need for a precise measurement procedure.

\subsection{Light yield, transparency and quenching}
\vspace{-0.1cm}

We used the light quenching model introduced in \cite{Escada}, with the 2- and 3-body quenching rate constants
measured by Wojciechowsky in \cite{Woj} for the Xenon triplet state in the presence of CH$_4$. In this simplified model, Xenon scintillation results from the 
competition between excimer formation and the quenching of atomic species. We realized that this model describes well earlier S$_1$ 
data from Pushkin \cite{Push} once corrected for recombination light. A detailed description of the simulation procedure
will be included in \cite{Aze}, which is currently under preparation: generally, when the 3-body quenching rate
constants were not available for some of the gases, an analogy with the systematics measured in \cite{Woj} was used. The gases chosen are all highly transparent to Xenon scintillation, with exception of \cod. For the \cod\ concentrations proposed here, however, transparency over 1 m is expected to comfortably exceed 50\% (see also \cite{Aze}).

The simulation toolkit used in this work for computing the light yields is an evolution of the one described in \citep{Oliveira:2011xx},  being now able to include molecular gas mixtures. For that, the existent GARFIELD++ code was changed in order to include the method described in \citep{Escada}.


%
%
%


The most sensitive parameter to both fluctuations of the light generation process and the light yield is the energy resolution. The FWHM energy resolution, $R_E$, of an electroluminescence-based  (EL) detector corresponds to the sum in quadrature of the contributions of the different processes occurring in the detector.  According to \cite{carlosres} the expression for the energy resolution can be written as:

\begin{equation}
\label{eq:res}
R_E = 2.35\sqrt{\frac{F}{\overline{N_e}}+\frac{Q}{\overline{N_e}}+\frac{2}{\overline{N_{ep}}}}
\end{equation}

where $F$ is the Fano factor, $Q$ represents the fluctuation associated with the light generation process and the last term reflects the variations in the number of photoelectrons extracted from the PMT photocathode per event. 
In order to minimize the energy resolution the two last terms in equation \ref{eq:res} should be as small as possible. On the other hand, the objective for the NEXT collaboration is to have an energy resolution between 0.5 and 1\% at \qbb.  That relaxes somewhat the conditions on the two last terms.

\begin{figure}[h!]
\centering
\includegraphics[angle=0, width=0.77\textwidth]{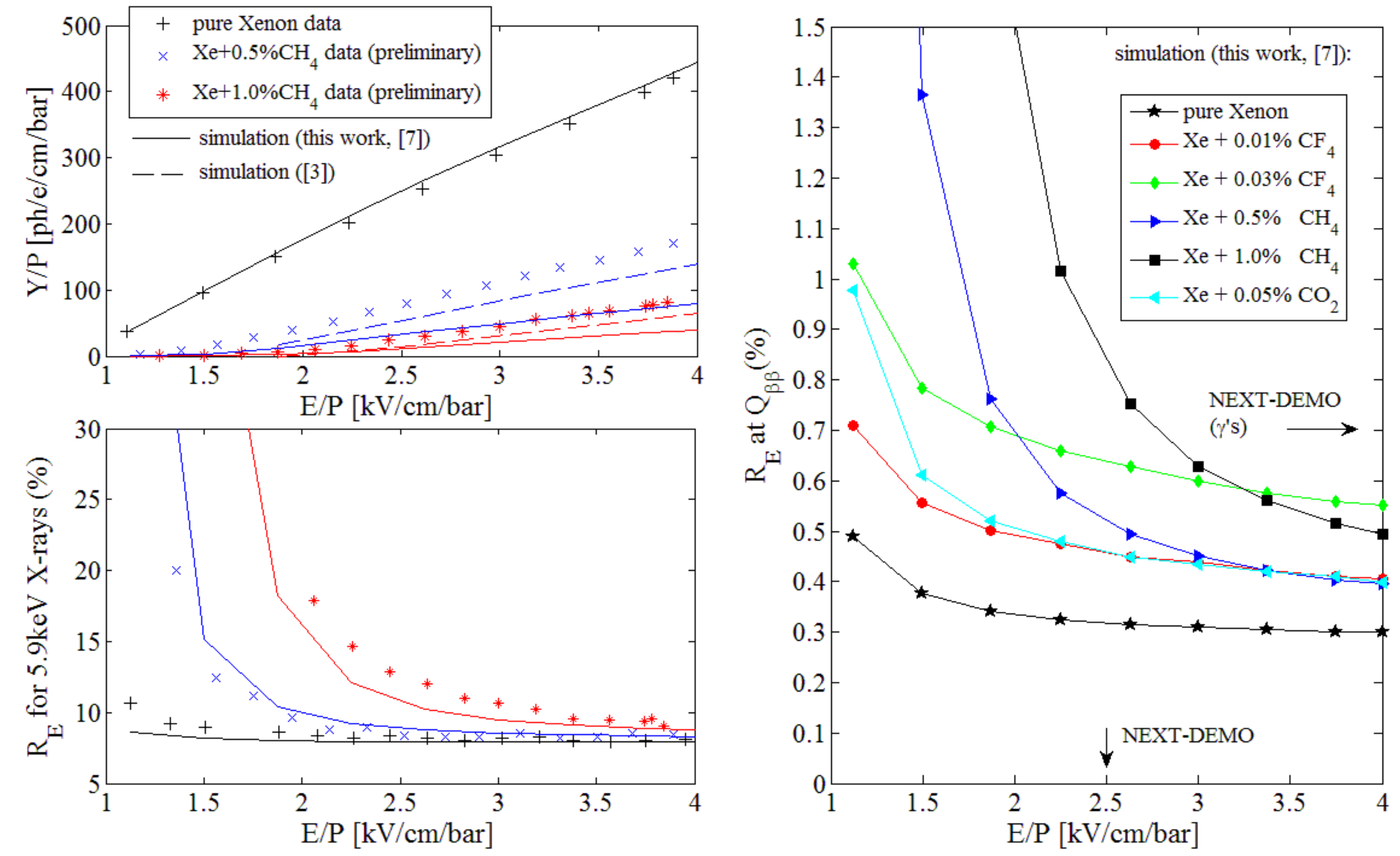}
\caption{Left-top: secondary light yields (EL) from simulation and comparison with preliminary data taken with a drift-less Gas Proportional Scintillation Counter (GPSC) \citep{simsancon2001} at 1bar.
Left-bottom: energy resolution in data (GPSC) and simulation. Simulations have been matched to the pure Xenon case by including
a flat 5.5\% instrumental contribution. A 30\% light collection efficiency was assumed.
Right: simulated energy resolution for the \qbb\ energy in a 5mm EL gap at 10 bar, assuming a light collection efficiency of 3\%. The simulation 
code has been introduced in the text and will be thoroughly described in \cite{Aze}. } \label{fig:eres}
\end{figure}
%

The graphs shown on the left in figure \ref{fig:eres} show the results from simulations and a comparison with preliminary data obtained in a small setup. During the measurements, there was no direct determination of the concentration of the molecular additive, which may differ slightly from the intended mixture due to the presence of the purifiers in the system. This could partially explain the imperfect agreement with results from simulations. An RGA will be installed in the system in order to clarify this aspect.

The graph on the right in figure \ref{fig:eres} shows the expected energy resolution for an electroluminescence-based  detector operating at different reduced fields. This graph shows the behavior of the different mixtures. On the one hand, when the electric field is too low the resolution worsens rapidly due to the fluctuations in photoelectron detection. On the other hand, the technical limitations of maintaining stable conditions in a large detector with a very high operating voltage in the EL region limits the realistic fields to values below 3.5kV/cm/bar. Still, figure \ref{fig:eres} shows a number of mixtures that will give an acceptable energy resolution while substantially modifying the electron diffusion.
A more detailed look at figure \ref{fig:eres} (right) shows curves with two different behaviours. For some of the mixtures the energy resolution stabilizes around similar values for relatively low reduced electric fields ($\sim$1.5 kV/cm/bar) while, on the other hand, for \chf\ mixtures, the energy resolution saturates much later and dramatically worsens when reducing the field. Those two different behaviors are clearly related to two different physical processes that affect the energy resolution. In the case of \chf\ mixtures the limiting factor is the light yield, which is the reason why high fields (high yields) are necessary to reach the stable region. Other mixtures stabilize much earlier, so this behaviour can not be related to the light yield but to the intrinsic fluctuations in the light generation process, which are dominated by the onset of dissociative attachment in both \cod\ and \cff.

\vspace{-0.3cm}
\section{Conclusions}
\vspace{-0.3cm}

A series of simulations on electron mobility and scintillation in gaseous xenon with different mixtures and concentrations of molecular gases has been performed. Simulations show that there are several molecular additives that can significantly reduce the electron diffusion in the detector down to the limit given by our instrumentation. Those mixtures can offer at the same time a good energy resolution, at the level of $0.5\%$ at \Qbb. 
The NEXT collaboration is making an R\&D effort along this promising line, using the small setups existing at various labs of NEXT member groups. Ultimately, the concept will be verified in NEXT-DEMO where measurements of diffusion, primary scintillation and energy resolution can be done simultaneously. This step will allow to assess if the small concentrations required can be maintained stable in a large experiment, even in the presence of purifiers.

%

\vspace{-0.3cm}

\acknowledgments
\vspace{-0.3cm}

This work was supported by the following agencies and institutions: the European Research
Council (ERC) under the Advanced Grant 339787-NEXT; the Ministerio de Economia y
Competitividad of Spain under grants CONSOLIDER-Ingenio 2010 CSD2008-0037 (CUP),
FPA2009-13697-C04 and FIS2012-37947-C04; the Portuguese FCT and FEDER through the
program COMPETE, project PTDC/FIS/103860/2008; and the Fermi National Accelerator
Laboratory under U.S. Department of Energy Contract No. DE-AC02-07CH11359.
C.D.R. Azevedo was supported by PostDoctoral grant from FCT (Lisbon) SFRH/BPD/79163/2011. C.M.B. Monteiro was supported by Grant from FCT (Lisbon) SFRH/BPD/76842/2011

The authors also want to acknowledge Steve Biagi for his interest and discussions.


\end{document}

%% file: defs.tex
\newcommand{\bbonu}{\ensuremath{\beta\beta0\nu}}

\newcommand{\qbb}{\ensuremath{Q_{\beta\beta}}}
\newcommand{\Qbb}{\ensuremath{Q_{\beta\beta}}}

\newcommand{\BI}{\ensuremath{{}^{214}}Bi}

\newcommand{\Xe}{\ensuremath{^{136}}Xe}

\newcommand{\Tl}{\ensuremath{^{208}}Tl}

\newcommand{\cod}{CO\ensuremath{_{2}}}
\newcommand{\chf}{CH\ensuremath{_4}}
\newcommand{\cff}{CF\ensuremath{_4}}


%% file: monrabal.bbl
\begin{thebibliography}{9}
\vspace{-0.3cm}

\bibitem{GomezCadenas:2011it} G\'omez-Cadenas, \textit{et. al.}, M., Riv.\ Nuovo Cim. 35(2012)29-98
\bibitem{Ferrario:2015kta}
  P.~Ferrario {\it et al.} [NEXT Collaboration],
  arXiv:1507.05902 [physics.ins-det].

\bibitem{Escada} J. Escada et al., JINST 6(2011)P08006.
\bibitem{Woj} K. Wojciechowsky, M. Forys, Rad. Phys. Chem. 54(1999)1. 
\bibitem{Push} K. N. Pushkin et al., Instrum. Exp. Tech. 49(2006)489.
\bibitem{simsancon2001} P. C. P. S.  Simoes \textit{et. al.}, X-Ray Spectrometry \textbf{30} (2001) 342-347.
\bibitem{Aze} C. Azevedo, D. Gonzalez-Diaz, et al., in preparation.

\bibitem{Oliveira:2011xx}
  C.~A.~B.~Oliveira {\it et al.},
  Phys.\ Lett.\ B {\bf 703} (2011) 217
  [arXiv:1103.6237 [physics.ins-det]].
\bibitem{carlosres}
Oliveira, C.~A.~B. \textit{et. al.}, \jinst{6}{2011}{P05007}
%


\end{thebibliography}
